# Towards Practical Privacy-Preserving Analytics for IoT and Cloud-Based Healthcare Systems

Sagar Sharma, Keke Chen, and Amit Sheth

Ohio Cneter of Excellence in Knowledge-enabled Computing (Kno.e.sis),
Wright State University, Dayton, OH 45435 USA
Email: *{sharma.74,keke.chen,amit.sheth}@wright.edu*

*Abstract—* Modern healthcare systems now rely on advanced computing methods and technologies, such as Internet of Things (IoT) devices and clouds, to collect and analyze personal health data at an unprecedented scale and depth. Patients, doctors, healthcare providers, and researchers depend on analytical models derived from such data sources to remotely monitor patients, early-diagnose diseases, and find personalized treatments and medications. However, without appropriate privacy protection, conducting data analytics becomes a source of a privacy nightmare. In this article, we present the research challenges in developing practical privacy-preserving analytics in healthcare information systems. The study is based on kHealth—a personalized digital healthcare information system that is being developed and tested for disease monitoring. We analyze the data and analytic requirements for the involved parties, identify the privacy assets, analyze existing privacy substrates, and discuss the potential tradeoff among privacy, efficiency, and model quality.

I. INTRODUCTION

A modern healthcare system is a complex data-driven mesh which relies on continuous patient monitoring, data streaming and sharing, and use of advanced big data analytics to provide essential health services to patients. It consumes information from patients' electronic health records (EHRs) including past diagnoses, hospital visits, interactions with the doctors, lab results (e.g., X-Rays, MRIs, and EEG results), past medications, treatment plans, and post-treatment complexities [11]. Furthermore, it consumes Internet-of-Things (IoT) sensor data that track and stream the patients' physical attributes such as activity level, heart rate, oxygen saturation, temperature, and respiration flow rates in a synchronous manner. A modern healthcare system stores and analyzes the collected data to build analytic models that provide myriad of health services to the patients, such as real-time monitoring for identifying health anomalies [9].

Easily available IoT devices and plethora of diverse datasets have opened the doors to building competitive personalized health analytics models able to detect abnormal change in one's health, predict the likelihood of a clinical event, and warn about the onset of conditions. The types of models and analytics involved vary from simple statistical aggregations to more complex data mining and machine learning, including natural language processing and even deep learning. Patients can now subscribe to the monitoring services or emergency detection platforms that use data generated models to detect health anomalies, suggest preventive measures or home remedies, and trigger hospital visits on the basis of severity.

The advantages of data-driven healthcare systems, however, come with a price, an extraordinary effort to protect patients' privacy without compromising the utility of the data and associated healthcare services. Like in education and social networking, protecting privacy in healthcare systems is urgent and challenging because of the sensitivity of the associated data and the multi-faceted exposure of the healthcare systems. Unauthorized exposure of the sensitive healthcare data not only violates Health Insurance Portability and Accountability Act (HIPAA), it can have lifelong social, economic, and health-related consequences to the patients. Doctors, nurses, lab technicians, emergency responders, and prescription handlers constitute the healthcare teams and have access to patients' data. Similarly, researchers and other model consumers require patients' data and the generated models for their scientific studies and commercial purposes, including some potentially against patients' best interests. In addition, the sheer amount of data and associated complexity of analytics require healthcare systems to employ third-party cloud infrastructure providers for massive-scale parallel processing and data storage. These exposures cannot be eliminated because of the critical roles the participating entities play.

**THE PRIVACY CHALLENGES.** Preserving data privacy from adversary parties in a healthcare information system without affecting data utility, model learning, and data sharing is challenging. An ideal privacy situation would require data and models always be protected outside the data contributors' or the users' personal devices. Privacy must be maintained throughout the storage, processing, and communication phases of an analytic task. Such an ideal privacy is almost unachievable. For instance, applying super-secure encryption schemes, such as AES-256 encryption, to protect data from potential adversaries seriously jeopardizes computations and services a typical healthcare service provider yearns to provide. On the other side, settling with something simpler like data anonymization or order preserving encryption proves to be inadequate against privacy breaches and information theft.

The selection of data protection method also affects the design of privacy-preserving analytics algorithms. As protection techniques only provide room for limited operations on the obfuscated data, complex algorithms must be disintegrated to these simpler operations. for example, a framework using homomorphic encryption to protect data must express the data mining algorithms in terms of simple homomorphic additions and multiplications.

Furthermore, privacy-preserving frameworks incur increased overall communication, storage, and computation costs, many times making them hapless in real-world scenarios. for example, even the cost-effective additive homomorphic encryption scheme, Paillier encryption, turns a floating-point plaintext value to a 256 byte ciphertext with a 1024-bit security key, and it takes about 5 seconds to sum 1,000 such encrypted ciphertexts. Thus, it becomes vital to distribute total workload of privacy- preserving frameworks to their participants relative to the resources available to them. A practical framework must ensure the resource-constrained parties perform lighter complexity tasks, while the expensive tasks be parallelized at the resource-abundant party such as a cloud.

**GOALS.** Using the context of kHealth, a mobile/digital health monitoring and management system currently being evaluated with patients and physicians, we elaborate on the privacy risks, features, and challenges of privacy-preserving analytics in IoT and cloud based healthcare information systems. We focus on the building of privacy-preserving data mining algorithms relevant to healthcare informatics and then analyze the candidate solutions. While discussing these solutions, we will try to understand their intrinsic tradeoffs amongst privacy, cost, and utility.

## II. IoT HEALTHCARE FRAMEWORK

We use an IoT-based health monitoring system, kHealth (http://knoesis.org/projects/kHealth) [2] to illustrate typical IoT healthcare system frameworks. kHealth uses both personal and physiological observations, sensed with wearable devices on subscribed patients as well as population (e.g., Twitter and a weather service) and public (e.g., CDC and hospital provided) level data, to generate personalized predictive models. The IoT sensors track and stream to the kHealth provider readings such as peak respiratory flow rate, weight, and activity level in addition to location and other environmental attributes (e.g., outdoor air quality index (AQI), pollen, humidity, mold, ozone, hydrocarbons, nitric oxide, carbon monoxide, and carbon dioxide levels). In addition to the sensor data, the kHealth provider exploits public datasets that provide statistics about the propensity of disease occurrences for various demographic and socioeconomic features. kHealth deploys machine learning and other data mining models, including Semantic Web technology to analyze and identify the status of a patient's condition and recommend cautions or alarm an immediate medical care. In summary, kHealth abstracts health signals from wearable devices and other various datasets, extracts relevant features, and builds personalized health predictive models for its subscribers and authorized researchers/doctors.

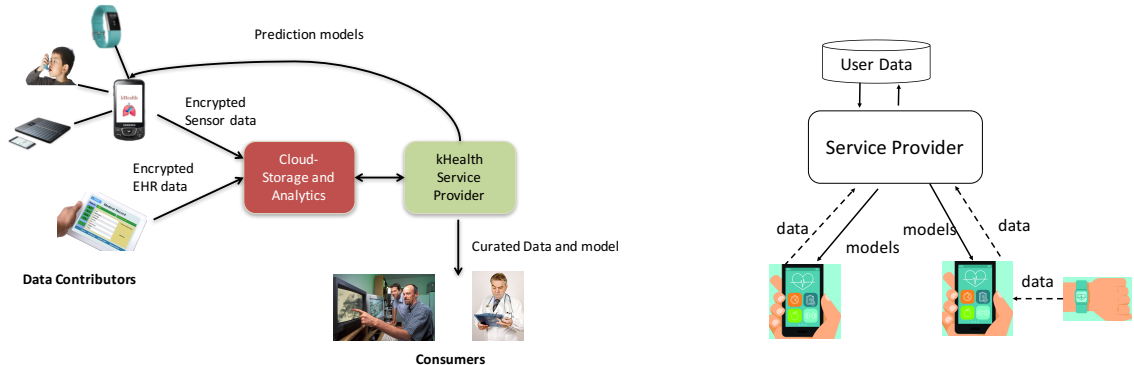

Figure. 1 **Left: The kHealth Framework for Asthma Management. Right*: Involved Parties and Interactions.*** Encrypted personal sensor data and EHRs, together with public and population level data, is collected and analyzed to build and distribute models that aid in asthma control and continuous monitoring for Asthma outbreaks. The kHealth service provider may use a cloud to process data. Personalized models are transferred to the patients' devices, and global models are shared by model consumers.

*A. INVOLVED PARTIES*

A system like kHealth involves several parties:

- the healthcare system provider (i.e., the service provider or SP),
- patients (i.e., the distributed data contributors or users),
- medical providers (doctors, hospital staff, and nurses),
- researchers (i.e., the model/data consumers), and
- the cloud on which the healthcare system provider may rely for data processing and storage.

Figure 1 shows the kHealth system, the involved parties and their interactions.

SPs store patients' data (encrypted, perturbed, or anonymized, and without any Personally Identifiable Information (PIIs)), and are interested in data mining and maintaining propriety of the collected data and learned models. An SP may choose to outsource data and computation to a cloud provider that delivers infrastructure for storage and analytics. The users or the patients are distributed individuals that subscribe to an SP's services. They may possess wearable sensors, healthcare monitoring devices, and smartphones acting as sensors or gateways to the SP. The data/model consumers in the framework are the doctors, health care providers, or researchers that avail from the data and insights brought by the generated models. The doctors are concerned with monitoring the patients and detecting any abnormalities remotely, while the researchers are driven to broader perspective, for example studying the prevalence of a symptom in a control group of patients or performing longitudinal studies. In some scenarios, the doctors and healthcare providers might also contribute patients' EHRs and lab assessments to be shared with the SP.

*B. ANALYTICS COMPONENTS*

Several data analytic methods are prevalent in digital health applications. From statistics to machine learning, there are important real-life applications of these data mining categories. The core of privacy-preserving analytics is to develop the privacy-preserving versions of these algorithms that work with different systems and security assumptions.

**Statistical Summarization.** Statistical summarization of an individual's data provides valuable knowledge about her medical conditions. Statistical measure of global, population, and personal observations can be both descriptive and inferential.

**Supervised Learning.** Automated classification, regression analysis, and neural networks are invaluable components of IoT-based healthcare systems. With domain-expert labeled examples, supervised learning can effective in diagnosis of illness, identification of disease-phenotype attributes from Genome-wide association studies, and predicting global outbreaks of diseases.

**Unsupervised Learning.** When data lacks class labels and is extremely large and therefore difficult to manually annotate, unsupervised learning is applied. These methods, such as clustering, semantic indexing, and dimensionality reduction, help identify the patterns and trends in healthcare data.

Analytic models learned by a service provider can be of global or personalized nature. For examples, a global model that predicts the outbreak of a disease based on collections of patient data from different regions and a personalized health alarm system that adapts to individual's unique health conditions.

### III. APPLICATION SCENARIOS AND PRIVACY ISSUES

An IoT and cloud based framework, as described in Section 2 must address several privacy issues and challenges. Depending on the application scenario and variations in the involved parties, their trustworthiness, and the constituting interactions, the privacy issues and challenges differ. In this section, we first discuss the private assets at risk and then examine two important application scenarios in typical IoT-based health informatics frameworks: 1) outsourced computations, and 2) information sharing. Outsourced computation relates to everything that happens to users' data outside the perimeter of their personal devices and home networks. It is vital to examine the extent of privacy exposure to the involved parties during storage, analytics, communication, and intermediate phases in the outsourced scenarios. On the other hand, information sharing enables interested parties to learn global models or data, or share their data to learn models together. It must be ensured that there is no breach to individual privacy while sharing data and models. We identify the privacy exposures, and discuss ways to mitigate them in both the categories.

**Privacy Assumptions.** Typically, when we consider the privacy issues only, we can assume that all adversarial parties are "honest-but-curious" - they perform their tasks obediently, i.e., guarantee data and model integrity and follow the provided protocols exactly; however, they might surreptitiously snoop for information to benefit from. We assume all infrastructures and communication channels are secured from external hacking, co-resident attacks, or active breaching with strong security means. Researches on infrastructures and communication security together with data and model integrity checking are orthogonal to this article's objectives.

## A. PRIVATE ASSETS AT RISK

**User Data.** An SP collects diverse data from the users including EHRs and sensor data. In the proposed kHealth framework, the collected data may be of personal and sensitive nature. Direct observation or inferences from attributes such as activity levels, location, respiratory flow rate, weight, and heart-rate may reveal an individual's health conditions, activity patterns, location traces, and life choices. Adversaries might be interested in the values of specific data records such as the feature attributes, the frequencies, and the group associations or the class labels.

**Generated Models.** An SP generates several statistical, data mining, and machine learning models to serve the needs of patients, the doctors, and itself. The global models must not be biased towards an individual or a certain group, i.e. they should not aid in distinguishing one individual from another. The personalized models must be delivered to the authorized individuals only, for example a patient's doctor, in a private manner. An adversary can examine a model and test it to probe into individual's private information. This is a critical problem when the adversary knows the type of model being generated or the approximate subjects of data collections. For example, a disease detection model based on subjects' physical, racial, and spatial attributes when leaked, allows an adversary to plug into the model individuals' visible attributes to infer her health conditions.

**Intermediate Results.** The intermediate results of model learning can reveal information about both the model and the training data. The adversary should not be able to gather any relevant user information from such results. Though knowing some of the intermediate results, e.g., the number of iterations, the convergence of a model, the size of a model, and the elapsed time might be essential for accurate model learning, care must be given so that no additional knowledge is revealed to an adversary than what is already known.

## B. OUTSOURCED COMPUTATION

Outsourced computation has become common with the development of cloud computing and popular use of mobile and IoT devices. Protecting privacy in outsourced computation involves two typical scenarios: outsourcing tasks by SPs to untrusted cloud providers for resource-intensive processing and resource-restricted end users outsourcing to untrusted SPs.

**Trusted Service Providers Outsourcing to an "Honest but Curious" Cloud Provider.** A fully trusted SP gathers subscribed patients' data, then analyzes the data to construct prediction models. No privacy risk is suspected from the SP, its associates or any third party involved. As a research product of a reputed university and government funding, we can safely establish kHealth as a trusted SP. Often, a trusted SP must depend on public cloud infrastructures for storage and computing elasticity. One immense advantage of such setup to the SP is that it can outsource heavier computations to the cloud provider without having to maintain expensive in-house infrastructures.

The SP needs to shield all the privacy assets from the cloud provider as the curious cloud may use the assets to derive its own models for its commercial gains. This mandates data to be encrypted or perturbed by the SP or its subscribing users before submission to the cloud. Meanwhile, introducing privacy to the framework should preserve the benefits of cloud computing: the heavier storage and higher complexity computations should be carried out by the cloud provider and the SP or the users must be minimally involved. Thus, frameworks like kHealth must ensure the cloud takes over the complex data mining algorithms working with encrypted or perturbed data while minimizing the patients' and the SP's involvement. An example of such a framework is presented in "Privacy-Preserving Spectral Analysis of Large Graphs in Public Clouds [14]. That paper proposes a framework that protects all privacy assets from the cloud provider while achieving the objective of outsourcing storage and conducting spectral analysis to a public cloud.

**End-Users Outsourcing to "Honest but Curious" Service Provider.** This is an extreme scenario where the users do not trust the SP itself, i.e., the SP might be commercially motivated or not credible enough to win users' faith. The users, however, are allured to the services the SP offers. When an SP is not a reputed organization or a government entity; for example, a commercial activity tracking analytics provider, it can be considered an honest but curious party. All the privacy assets must be protected from such an SP and any third-party it collaborates with, including the cloud. The privacy constraints obligate the SP to provide its users with mechanism to submit encrypted or protected data and develop algorithms that generate analytic models using the obfuscated datasets.

This is a particularly tricky situation as complex data analytics become very expensive, if not impossible, over encrypted data without intermediate decryption or unmasking. A possible remedy is to introduce another honest-but-curious party, called crypto-service provider (CSP), which will manage secret keys, decrypt intermediate results, and assist SP to finish the modeling task. The CSP's workload should be appropriate (i.e., the heavy processing must be done by the cloud or the SP). Most importantly, CSP must be held accountable to the end users and cannot collude with the SP. Researchers have proposed CSP- based frameworks to build privacy-preserving ridge regression and matrix factorization [12]. Figure 2 depicts a framework where SP and CSP learn models over encrypted/masked data and the generated models are only decodable by the individual users.

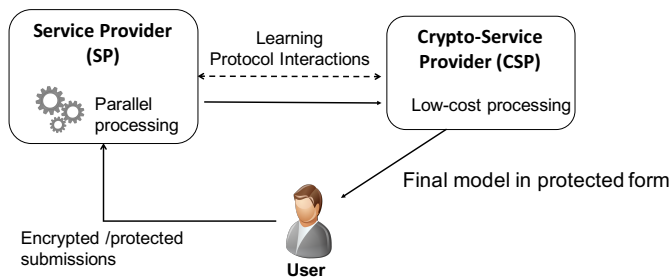

Figure. 2 Predictive models are collaboratively generated by honest-but-curious cryptographic service provider (CSP) and SP with preserved privacy.

## C. INFORMATION SHARING

Healthcare analytics and data must be shared time to time with both trusted and semi-trusted parties, including researchers, medical providers, insurance companies, pharmacies, lab technicians, etc. to allow them to learn global trends or models without breaching individual privacy. For example, in the kHealth framework, researchers wish to access the users' datasets to analyze global patterns or test new algorithms. The kHealth provider must take care of the individual privacy when sharing the private data with the model consumers or other health service providers.

Broadly, information sharing happens in three ways:

a) An interactive manner. An SP provides a query interface to the data consumers who submit statistical queries,

b) A non-interactive manner. An SP publishes curated datasets with users' identities and virtual identifiers removed for researchers/consumers to explore, and

c) A collaborative manner. Multiple parties (for e.g. several SPs) collaboratively generate models from combined/shared data without revealing raw private data.

The interactive setting is considered easier to apply privacy walls as an SP may strictly define types and numbers of queries allowed through its interfaces via mechanisms like differential privacy [6]. The non-interactive setting opens doors to more severe form of privacy leakage, as there is no control on types of attacks and model consumers can learn on the public datasets. The collaborative model learning by multiple parties, for example two or more hospitals, from union of their datasets without revealing the data to one another is addressed by the secure multiparty computation (SMC) using protocols such as Shamir's secret sharing [13].

## IV. CANDIDATE PRIVACY-PRESERVING BUILDING BLOCKS AND TRADE-OFFS

In this section, we outline the fundamental building blocks of privacy preserving frameworks and their trade-offs amongst multiple factors. Any healthcare informatics platform that fall under the application scenarios that we described in Section 3 needs to use one or more of these building blocks to ensure privacy-preserving analytics. It is crucial to familiarize these methods and comprehend their advantages and disadvantages. A fine balance amongst the tradeoffs of cost, privacy, and data utility may be possible when these methods are well understood and applied to the privacy-preserving analytic frameworks.

### A. METHODS AND TRADE-OFFS FOR OUTSOURCED COMPUTATION.

**Expressive but Expensive Methods.** The two well-known generic approaches for privacy-preserving computation on untrusted platforms are fully homomorphic encryption (FHE), and Yao's garbled circuits (GC) together with Oblivious Transfer (OT). FHE allows arbitrary number of additions and multiplications on encrypted data without decryption. GC provides basic logic gates such as AND, XOR, and OR, with which more complex circuits can be implemented. Conceptually, these techniques can be used to construct any data mining algorithms. However, the schemes are very expensive in storage, communication, and computation. The ciphertext resulting from the FHE encryption schemes [4] becomes several magnitudes larger than plain text data, placing an excessive burden on storage and communication. Likewise, the best optimized GC implementation incurs impractical communication costs between the participating parties [12]. Moreover, building and implementing complex algorithms with GC circuits is a challenging task.

**Efficient but Less Expressive Methods.** If both the highest privacy preservation (e.g., semantic security) and practicality are desired, the best hope is the use of somewhat homomorphic encryption (SHE) and additive homomorphic encryption (AHE) schemes. The popular SHE schemes include ring learning with error (RLWE) [4] and Pairing (BGN) [3] schemes, that allow an arbitrary number of homomorphic additions with one [3] or a small number of multiplications (e.g., <5) [4]. This limitation restricts the types of data mining algorithms that can apply SHE schemes to adapt to the privacy standards. Moreover, RLWE cryptosystem suffers from heavier cipher-size which means heavier communication and storage cost. Though efficient methods for message packing exist, the cost of manipulating the packed messages is still high. Despite the challenges and limitations, RLWE scheme has been used in the ML Confidential method for classifier learning [8]. Pairing(BGN) scheme on the other hand, suffers from unacceptable decryption performance for larger integers (16 or 32 bits) making it impractical in analytics requiring high precision and involving large datasets.

The Paillier cryptosystem is a popular AHE scheme, efficient in encryption, decryption, and homomorphic computation, and its ciphersize is much smaller than RLWE scheme. However, it requires one of the operands in homomorphic multiplication to be unencrypted, which causes privacy leakage. Hence, the use of any AHE scheme must be complemented with methods to protect the unencrypted operand. A good example is the PrivateGraph [14] approach designed for privately analyzing outsourced graph data in the cloud. It uses Paillier cryptosystem to encrypt data and protects the unencrypted operands in homomorphic multiplications with a novel noise generation and an efficient noise-recovery mechanism. The approach achieves a practical work allocation between the cloud and the data owner - the cloud takes over the $O(N^2)$ operations that can be further improved with parallel processing, while the data owner's cost is $O(N)$.

**Efficient Methods with Weaker Privacy Notion.** If one can settle with weaker security notions, data perturbation methods [5], [15] and order preserving encryption (OPE) schemes [1] are the best choices. Data perturbation techniques such as additive perturbation, rotation perturbation, random projection perturbation, geometric perturbation [5], and random space perturbation (RASP) [15] transform the entire datasets with noise injection for outsourcing. Unlike the homomorphic encryption schemes discussed earlier, data perturbation normally applies mathematical transformations for floating-point values directly, and thus are much more efficient. Some techniques such as geometric perturbation [5] are extremely versatile as many existing data mining algorithms, without any modifications, can be applied to the perturbed datasets to learn models. Despite their practicality and ease of use, data perturbation techniques are vulnerable when attackers identify from external sources the original data distributions. One exception is the RASP mechanism which proves to be resilient to distributional attacks as well [15]. OPE methods preserve the ordering relationship among encrypted values enabling indexing of encrypted values. However, it also assumes adversaries do not know the original data distribution.

*B. METHODS AND TRADE-OFFS FOR INFORMATION SHARING*

So far, the most accepted scheme for information sharing is differential privacy for its rigorous theoretical foundation. Differential privacy [6] requires perturbing the outputs of sensitive functions, which affects the quality of models learned from the functional outputs. A trade-off between model quality and privacy may exist if the models are sensitive to the perturbation. Therefore, differential privacy may not be ideal for applications requiring the highest-quality models. In addition, it is difficult to satisfy differential privacy in a non-interactive setting [10]. Data anonymization methods [7] have been used for micro- data publishing (i.e., the non-interactive setting). However, without rigorous theoretical foundation, they suffer from various background-knowledge based attacks. Shamir's secret sharing[13] allows dispersion of secrets amongst several parties such that the secrets can be revealed only when a threshold number of parties agree. The parties can also partake in a secure multiparty computation(SMC) interaction to generate collaborative models with the shared secrets without revealing the underlying raw data. One disadvantage of this scheme is that all the participants need to intensively involve in the computation.

## V. CONCLUSION

In this article, we dissected the state of privacy in modern IoT-based healthcare systems that rely on patients' data to produce predictive models beneficial to the patients, medical providers, and researchers. A straightforward solution that guarantees privacy together with practical and quality outcomes is difficult to blend. First, a careful analysis of the involved parties, the privacy assets at risk, the desired algorithms/models and model quality, and the audience of the models must be conducted. Depending on the desired analytics and privacy level, an additional party such as a cryptographic service provider might need to be introduced to a framework. Privacy primitives such as homomorphic encryption schemes, data perturbation, and differential privacy may be applied in the healthcare frameworks after ensuring the correctness and balancing the associated cost of computation, communication, and storage. It might not be possible to implement the best healthcare modeling algorithms in the privacy standard because of numerous tradeoffs and restrictions. However, the drawbacks must be compensated with parallelizable methods such as ensemble learning to generate reliable models. Modern healthcare frameworks like kHealth can overcome privacy challenges and yet be practical if right privacy building blocks and the applicable scenarios are identified and implemented. When the privacy and practicality aspects of privacy-preserving healthcare informatics are balanced, we will see development of fruitful healthcare applications and their fast and effective adaptation.

BIOGRAPHIES

**Sagar Sharma** is currently a PhD student in the Department of Computer Science and Engineering at Wright State University. His research interests include privacy-preserving outsourced data mining, big data, cloud computing, and Internet of Things. (http://knoesis.org/students/sharma/)

**Keke Chen** is an associate professor in the Department of Computer Science and Engineering and directs the Data Intensive Analysis and Computing (DIAC) Lab of the Ohio Center of Excellence in Knowledge-Enabled Computing (the Kno.e.sis Center) at Wright State University. He earned his Ph.D. degree in Computer Science from Georgia Institute of Technology in 2006. His current research areas include secure data services and mining of outsourced data, privacy issues of social computing, cloud computing, Internet of things, healthcare informatics, and big data. During 2006-2008, he was a senior research scientist at Yahoo! Labs, working on web search ranking, cross-domain ranking, and web-scale data mining. He owns three patents for his work in Yahoo! Labs. (http://cecs.wright.edu/~keke.chen/)

**Amit Sheth** is the LexisNexis Ohio Eminent Scholar, the executive director of Kno.e.sis-Ohio Center of Excellent in Knowledge-enabled Computing, and an IEEE Fellow. His research interests are in Computing for Human Experience, Sematic-Cognitive-Perceptual Computing, Physical Cyber Social Big and Smart Data, Web 3.0, SmartIoT, and Semantic Web. He is one of the highly cited authors in Computer Science and has (co)founded four companies. (http://knoesis.org/amit/)